\documentclass[conference]{IEEEtran}
\IEEEoverridecommandlockouts

\usepackage{cite}
\usepackage{amsmath,amssymb,amsfonts}
\usepackage{algorithmic}
\usepackage{graphicx}
\usepackage{textcomp}

\usepackage{url}

\usepackage{fancyvrb}
\usepackage{xspace}
\usepackage{enumitem}
\usepackage[utf8]{inputenc}

\usepackage{mdframed}

\usepackage{soul}
\usepackage{graphicx}
\usepackage{balance}

\usepackage{colortbl}
\usepackage{xcolor}

\usepackage{multirow}

\newcommand{\caret}{$^{\wedge}$}
\newcommand{\tildes}{$\sim$}

\newcommand{\etal}{et al.\xspace}
\newcommand{\ie}{i.e.,\xspace}
\newcommand{\eg}{e.g.,\xspace}
\newcommand{\fig}[1]{Figure~\ref{#1}}
\newcommand{\tab}[1]{Table~\ref{#1}}
\newcommand{\sect}[1]{Section~\ref{#1}}

\newcommand{\maven}{\textsf{Maven}\xspace}
\newcommand{\cran}{\textsf{CRAN}\xspace}

\newcommand{\rubygems}{\textsf{RubyGems}\xspace}

\newcommand{\eclipse}{\textsf{Eclipse}\xspace}
\newcommand{\npm}{\textsf{npm}\xspace}
\newcommand{\npmsio}{\textsf{npms}\xspace}
\newcommand{\github}{\textsf{GitHub}\xspace}

\newcommand{\figsize}{0.9\columnwidth}
\usepackage{todonotes}

\newenvironment{custombox}{\begin{mdframed}[nobreak=false, innerleftmargin=4pt, innerrightmargin=4pt, innertopmargin=2.5pt, innerbottommargin=2.5pt]}{\end{mdframed}}

\newcommand{\ordd}[0]{{\mathrm{t}}}
\newcommand{\ordv}[0]{{\mathrm{v}}}

\newcommand{\func}[1]{{\mathsf{\mathbf{#1}}}}
\newcommand{\releases}[1]{\func{releases}(#1)}
\newcommand{\comp}[1]{{\mathit{#1}}}
\newcommand{\att}[1]{{_\comp{#1}}}

\newcommand{\prevt}[1]{\comp{prev}(#1)}
\newcommand{\nextt}[1]{\comp{next}(#1)}
\newcommand{\prevv}[1]{\comp{prev}_\ordv(#1)}
\newcommand{\nextv}[1]{\comp{next}_\ordv(#1)}

\newcommand{\satisfies}[0]{\vDash}
\newcommand{\pipe}[0]{~|~}

\newcommand{\available}[0]{\func{available}}
\newcommand{\installable}[0]{\func{installable}}
\newcommand{\missed}[0]{\func{missed}}

\newcommand{\deltat}[0]{\Delta_t}

\def\BibTeX{{\rm B\kern-.05em{\sc i\kern-.025em b}\kern-.08em
    T\kern-.1667em\lower.7ex\hbox{E}\kern-.125emX}}

\begin{document}

\title{On the evolution of technical lag in\\ the npm package dependency network
}
\author{\IEEEauthorblockN{Alexandre Decan, Tom Mens, Eleni Constantinou}
 \textit{Software Engineering Lab},
 \textit{University of Mons}\\
Mons, Belgium
}

\maketitle


\begin{abstract}
Software packages developed and distributed through package managers extensively depend on other packages.
These dependencies are regularly updated, for example to add new features, resolve bugs or fix security issues.
In order to take full advantage of the benefits of this type of reuse, developers should keep their dependencies up to date by relying on the latest releases.
In practice, however, this is not always possible, and packages lag behind with respect to the latest version of their dependencies. This phenomenon is described as technical lag in the literature.
In this paper, we perform an empirical study of technical lag in the \npm dependency network by investigating its evolution for over 1.4M releases of 120K packages and 8M dependencies between these releases.
We explore how 
technical lag increases over time, taking into account the release type 
and the use of package dependency constraints. We also discuss how technical lag can be reduced by relying on the semantic versioning policy.
\end{abstract}




\section{Introduction}
\label{sec:introduction}


Software developers are continuously confronted with the delicate choice between keeping their software in a stable, working state, and keeping it ``as up-to-date as reasonable" w.r.t. external dependencies (e.g., libraries or external systems) in order to benefit from bug fixes, security fixes, and relevant new features. However, updating to more recent versions might lead to an increased risk of backward incompatible changes.
Package maintainers advocate to investigate the impact of breaking changes in dependent packages before deciding to update them\footnote{``Not all scenarios will require you to update a packages as it could introduce breaking changes to your projects. Do the research first.'' (\url{https://www.thepolyglotdeveloper.com/2015/03/check-update-outdated-npm-packages/})}.
Depending on the number of dependencies, this can become an infeasible task.
To capture this challenging balance between risks and opportunities of updating dependencies, Gonzalez-Barahona et al.~\cite{Barahona2017OSS} proposed the concept of \emph{technical lag} for reflecting how outdated a software system is with respect to its upstream dependencies.
Having a precise way of measuring technical lag allows software developers to make informed decisions on whether and when to update their outdated dependencies.


Technical lag can be particularly important in software package distributions, where packages depend on each other to use third-party functionality and facilitate the development process~\cite{Zerouali2018,Kula2018EMSE}.
Dependencies between packages are defined based on dependency constraints, which specify the version range that is allowed to be installed at any given point in time.
Based on such constraints and a set of allowed installable versions, the latest allowed version of the required package is installed and used by a dependent package.
%
Package dependency networks of software package managers have been shown to be brittle because of the large and increasing number of dependencies over time~\cite{Bogart2016}. This is especially the case for the \npm package distribution, witnessing an exponential growth in number of packages and dependencies~\cite{DecanEMSE2018}.
Zerouali et al. provided preliminary evidence of technical lag in \npm by analysing package dependencies on a yearly basis~\cite{Zerouali2018}. 

We go one step further, by analysing the continuous evolution of technical lag at the level of package dependencies and package releases, and relating it to the release type of both dependent and required packages.
To this end, we focus on the following research questions:
\begin{itemize}[topsep=0pt]
\item $RQ_1$ How many packages have technical lag?
\item $RQ_2$ How long is the technical lag?
\item $RQ_3$ How frequently are packages updated?
\item $RQ_4$ When does technical lag increase?
\item $RQ_5$ When does technical lag decrease?
\item $RQ_6$ How could technical lag be reduced by proper use of semantic versioning?
\end{itemize}

By answering these questions, we aim to get a better understanding of technical lag throughout software packages in the \npm distributions.
These insights can help to bette manage and control technical lag, through tools to monitor and support package dependencies as well as through a more optimal usage of semantic versioning.

The remainder of this paper is structured as follows. \sect{sec:methodology} motivates the choice of \npm as a case study, introduces the necessary terminology and research methodology and describes the dataset used. Sections~\ref{sec:rq1}-\ref{sec:rq6} answer the research questions and \sect{sec:discussion} discusses our findings.
\sect{sec:threatsthreat} explains the threats to validity of our work and \sect{sec:related} presents related research.
\sect{sec:future} discusses the future work and \sect{sec:conclusion} concludes the paper.


\section{Methodology and Terminology}
\label{sec:methodology}


According to the 2018 Stack Overflow Developer Survey\footnote{\url{https://insights.stackoverflow.com/survey/2018}} to which over 100,000 developers participated, JavaScript is by far the most commonly used programming language (accounting for 69.8\%).
In addition, the \npm distribution 
was observed to have a higher distribution of package dependencies than other package distributions~\cite{DecanEMSE2018}. This increases the risk of packages suffering from technical lag due to outdated dependencies.
For these reasons we selected \npm for our empirical study.
The metadata of each \npm package release (such as the name, version, and list of dependencies) is stored in a \textsf{.json} manifest file.
Dependencies are used to specify other packages that are explicitly required by the release. The range of allowed versions can be restricted using dependency constraints.
By default, when a package is installed using the \npm package manager, the latest release of each required package satisfying the dependency constraint is installed.




Let us present the terminology and notations that will be used in the remainder of this paper.

\smallskip\noindent\textbf{Release.}
Let $E$ be a package distribution, \ie a set of packages.
Given a package $p\in E$, $\releases p$ denotes the set of releases of $p$.
Every release $r\in \releases p$ has a release date $r\att{date}$ and a version $r\att{version} = (\comp{major}, \comp{minor}, \comp{patch})$.
The triple $r\att{version}$ reflects the \emph{semantic versioning} mechanism suggested by \npm. A simple set of rules dictate how version components should be incremented when a package is updated. Package updates corresponding to bug fixes that do not affect the API should only increment the \textit{major} version component, backward compatible updates should increment the \textit{minor} component, and backward incompatible updates have to increment the \textit{major} component.

\smallskip\noindent\textbf{Release order.}
For any package $p$ we assume two total orders $<_\ordd$ and $<_\ordv$ on its releases, respectively based on their $\comp{date}$ and their $\comp{version}$. In the latter case, the versions are compared first based on their $\comp{major}$ component, then on their $\comp{minor}$ component, and then on their $\comp{patch}$ component.
For any release $r$ we write $\prevt r$ and $\nextt r$ to refer to the previous and next release w.r.t. $<_\ordd$, if it exists.
Similarly, we write $\prevv r$ and $\nextv r$ to refer to the previous and next release w.r.t. $<_\ordv$, if it exists.

\smallskip\noindent\textbf{Release type.}
For each $r \in \releases p$ such that $\prevv r$ exists, we define its release type $r\att{type}$ as \textsc{major} if $r_{version}$ and ${\prevv r}_{version}$ have distinct $\comp{major}$ components, as \textsc{minor} if they have similar $\comp{major}$ but distinct $\comp{minor}$ components, and as \textsc{patch} if they agree on all components except $\comp{patch}$.

\smallskip\noindent\textbf{Dependency.}
A release $r$ has a (potentially empty) set $r\att{deps}$ of dependencies. A dependency $d\in r\att{deps}$ is defined as a pair $(d\att{target}$, $d\att{constraint})$ composed of a target package $d\att{target} \in E$ and a dependency constraint $d\att{constraint}$ over the releases in $\releases{d\att{target}}$.
If $r' \in \releases{d\att{target}}$, we note $r'\satisfies d$ if $r'\att{version}$ satisfies $d\att{constraint}$.
Dependency constraints allow package maintainers to explicitly select the desirable or allowed versions of a dependency, and exclude the undesirable ones, e.g., those that can contain backward incompatible changes. Dependency constraints are typically used to specify a minimal (\eg $>=\text{1.2.3}$), maximal (\eg $<\text{1.3.0}$) or exact version (\eg $=\text{1.4.2}$) of a dependency.
The \npm package manager relies on the \textsf{semver}\footnote{\url{https://docs.npmjs.com/misc/semver}} tool to identify the version numbers satisfying a dependency constraint.
It provides specific version constraint notations, such as tilde \tildes~and caret \caret~to allow releases to benefit from backward compatible dependency updates. \tildes~allows for automatic updates of patches only, while \caret~allows for automatic updates of both patches and minor releases.
By default, the \npm package manager selects for installation the highest available version that satisfies the dependency constraint.

\smallskip\noindent\textbf{Available and installable releases.} Let package $p \in E$ and $t$ a point in time. The set $\available(p,t) = \{r\pipe r\in \releases p \land r\att{date}\leq t\}$ contains all releases of $p$ that were available for installation at time $t$.
Given a dependency $d$, the set $\installable(d, t) = \{r\pipe r\in\available(d\att{target}, t) \land r\satisfies d\}$ contains all available releases of the target package $d\att{target}$ that satisfy the dependency constraint.

\smallskip\noindent\textbf{Technical lag.}
Let $d$ be a dependency and $t$ a point in time.
We define $\missed(d, t) = \{ r\pipe r\in\available(d\att{target}, t) \land r >_\ordv \text{max}_{<_\ordv} \installable(d, t)\}$.
This set captures the highest releases (w.r.t. $<_\ordv$) of the target package that cannot be installed at time $t$ because of the dependency constraint. Only the releases that are higher than any installable one are comprised in this set.
We define the \emph{technical lag} $\deltat(d, t)$ induced by $d$ at time $t$ as follows:
\begin{align*}
\deltat(d, t)  =  \begin{cases}
t - \text{min}\{r\att{date}\pipe r\in\missed(d, t)\}  \\
0, \text{if $\missed(d, t)$ is empty}\\
\end{cases}
\end{align*}

\noindent Technical lag represents the time during which $d$ prevents the use of a newer version of its target package. By abuse of notation, we also use $\deltat(r,t)$ to refer to the technical lag of a release $r$ at time $t$, which is defined as the maximal technical lag induced by its dependencies:
\begin{align*}
\deltat(r,t) = \text{max}\{\deltat(d, t)\pipe d\in r\att{deps}\}
\end{align*}



\begin{figure}[!ht]
  \centering
  \includegraphics[width=\figsize]{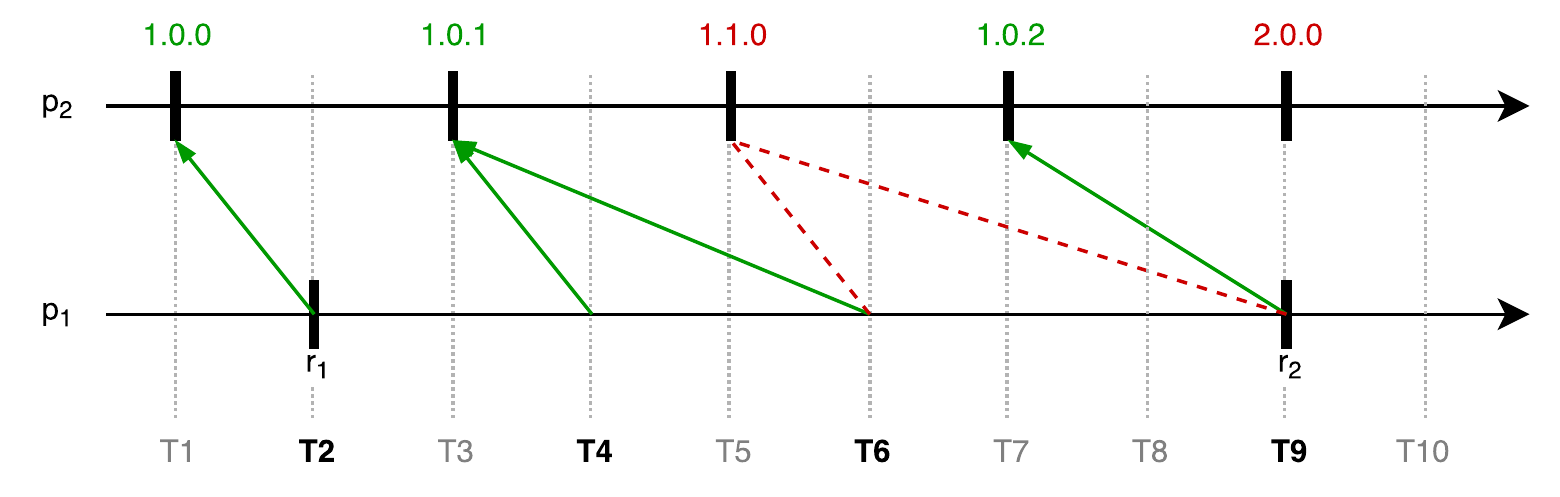}
  \caption{
    Selected release at $T_2$, $T_4$, $T_6$ and $T_9$ for dependency $(p_2, \text{\tildes1.0.0})$.
  }
  \label{fig:example_defs}
\end{figure}

\textbf{Example.} The following example illustrates our definitions.
Consider two fictitious packages $p_1$ and $p_2$.
Let $r_1\in\releases{p_1}$, such that $r_1$ has a dependency $d=(p_2, \text{\tildes1.0.0})$.
Constraint \tildes1.0.0 only allows patch updates over release 1.0.0.
Because of this dependency, the installation of $r_1$ requires a release of $p_2$ to be installed.
By default, \npm selects the highest available version that satisfies the dependency constraint. As a consequence, the selected releases of $p_2$ can be different depending on the installation time of $r_1$.
\fig{fig:example_defs} shows, by means of green arrows, which release of $p_2$ will be installed at different points in time $T_2$, $T_4$, $T_6$ and $T_9$.

\begin{table}[!ht]
  \caption{Installable and missed releases for dependency $d = (p_2, \text{\tildes1.0.0})$, and associated technical lag.}
  \label{tab:example}\centering

  \begin{tabular}{c|ccc}
    $t$ & $\text{max}_{<_\ordv}\installable(d, t)$ & $\missed(d, t)$ & $\deltat(d,t)$\\
    \hline
    $T_2$ & 1.0.0 & $\emptyset$ & 0 \\
    $T_4$ & 1.0.1 & $\emptyset$ & 0 \\
    $T_6$ & 1.0.1 & $\{1.1.0\}$ & $T_6 - T_5$ \\
    $T_9$ & 1.0.2 & $\{1.1.0, 2.0.0\}$ & $T_9 - T_5$ \\
  \end{tabular}

\end{table}

\tab{tab:example} shows, for each considered time point (first column), the release of $p_2$ that is selected during $r_1$ installation (second column), the set of releases of $p_2$ that are missed due to the dependency constraint (third column), and the resulting technical lag (fourth column).
For example, even though version 1.1.0 of $p_2$ is available at $T_6$, it will not be selected (indicated by the red dotted line in \fig{fig:example_defs}) because it does not satisfy the dependency constraint \tildes1.0.0.
Consequently, $r_1$ does not rely on the highest available version of $p_2$ at $T_6$ and has a technical lag induced by its dependency $d$.
This technical lag is computed as $\deltat(d, T_6) = T_6 - T_5$, the difference between the considered time point $T_6$ and the release date $T_5$ of the missed version 1.1.0.
At $T_9$, even if the latest version 1.0.2 of $p_2$ satisfies the dependency constraint \tildes1.0.0, release $r_1$ is still lagging behind because the higher version 1.1.0 is missed.
The technical lag of $d$ at $T_9$ is then $\deltat(d, T_9) = T_9 - T_5$.

Assume that a new version $r_2$ of $p_1$ is released at $T_9$ with a dependency constraint $d' = (p_2, \text{\caret1.0.0})$ allowing both minor and patch updates starting from version 1.0.0 of $p_2$.
At $T_9$, the highest installable version of $p_2$ that satisfies this constraint is 1.1.0.
Even if version 2.0.0 is already ``missed'', $\deltat(d', T_9) = 0$ because it corresponds to the difference between the considered time $T_9$, and the release time of the ``missed'' version 2.0.0, \ie $T_9 - T_9 = 0$.
At $T_{10}$, however, we have $\deltat(d', T_{10}) = T_{10} - T_9$ that reflects that version 2.0.0 is missed for a certain amount of time.


\textbf{Dataset.} Our analysis relies on the 2017-11-02 dump of the open source discovery service \textsf{libraries.io}~\cite{andrew_nesbitt_2017_808273}. 
Since we focus on \npm packages, we only consider the metadata from the manifest of each package provided by the official \npm registry.
For each release of each \npm package, we consider its list of dependencies. We restrict ourselves to runtime dependencies only, since we are only interested in those dependencies that are required to install and execute the package.
We also exclude dependencies that target packages that were not available through the package distribution (e.g., packages that are hosted directly on the web or on Git repositories).
This pre-filtering step resulted in 610,096 \npm packages, 4,202,099 releases of these packages and 20,240,402 runtime dependencies between them.


We applied additional filters to our dataset for the technical lag analyses.
First, we excluded 322,840 pre-releases of packages, based on their version number, \eg 1.0.0-alpha, 1.2.3-beta.0, 2.0.1-rc.
According to the \npm semantic versioning tool, such releases ``are meant to be unstable and are expected to have breaking changes'' and thus, \npm does not install a pre-release unless explicitly stated in the dependency constraint.
Next, we filtered out 427,568 packages (and their corresponding releases) that were either never updated (\ie packages that have only one release), not updated recently (\ie no update since January 2017), or isolated in the sense that they had no direct nor reverse dependencies and are thus of no interest in our analysis.

Our filtered dataset consists of 120,084 \npm packages, 1,447,709 releases of these packages and 8,044,034 runtime dependencies between them.
This represents 20\% of all packages, 35\% of all releases and 40\% of all dependencies of the original dataset.
The earliest package release date of the filtered dataset was registered on 2010-11-09, and the latest on 2017-11-02.
A replication package of our analysis is available on \url{https://doi.org/10.5281/zenodo.1283203}.

\section{$RQ_1$: How many packages have technical lag?}
\label{sec:rq1}
With the first research question we aim to gain an initial understanding of the omnipresence of technical lag in the \npm ecosystem.
To this extent, we quantify the technical lag of each release and each dependency over time.

For each package $p$ in \npm we gathered each release $r\in releases(p)$. 
At release date $r_{date}$ we computed the technical lag $\deltat(r, r\att{date})$ and $\deltat(d, r\att{date})$ for each dependency $d \in r_{deps}$.
Similarly, at the date of the next release (if any) we computed $\deltat(r, \nextt r\att{date})$ and $\deltat(d, \nextt r\att{date})$.
The difference between the two $\deltat$ values represents the number of dependencies or releases for which a technical lag got introduced somewhere during the life of the release. 

%

\begin{figure}[!ht]
   \centering
   \includegraphics[width=\figsize]{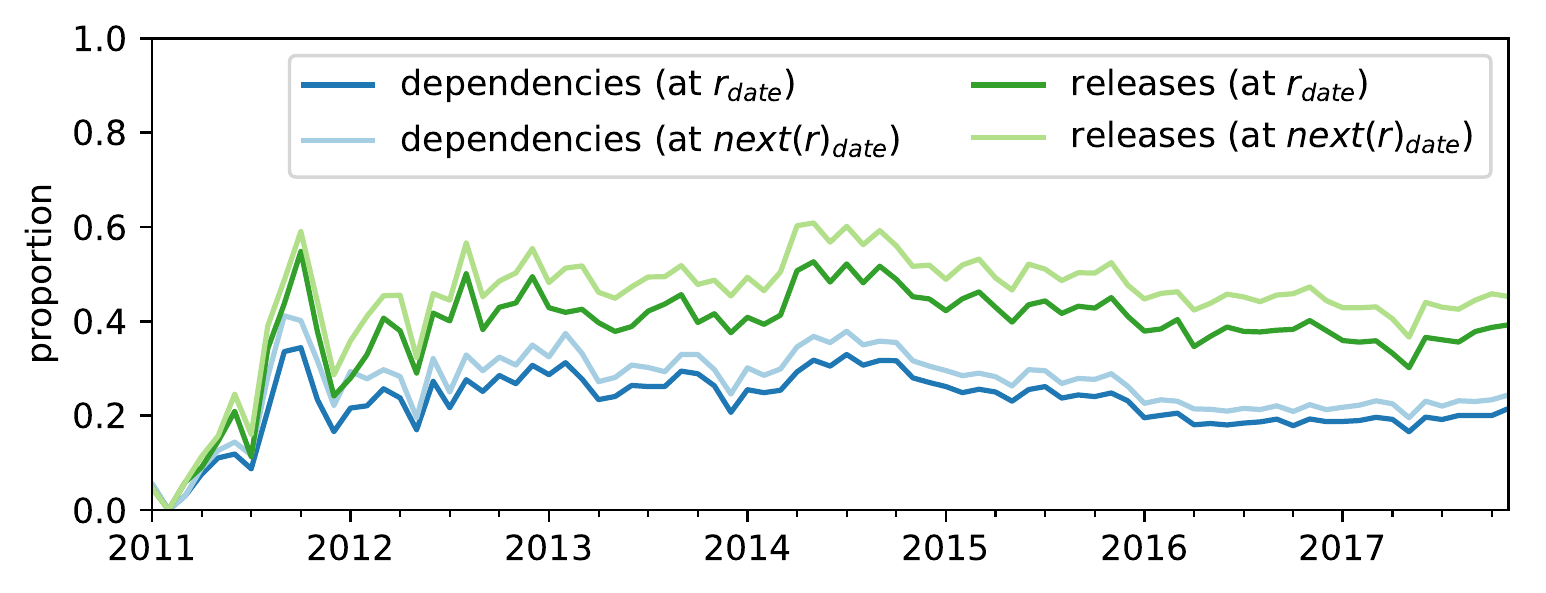}
   \caption{
     Monthly proportion of releases $r$ and dependencies $d\in r\att{deps}$ having a technical lag, \ie having $\deltat(r, t) > 0$ and $\deltat(d, t) > 0$ for $t\in\{r\att{date}, \nextt r\att{date}\}$.
   }
   \label{fig:temporal_proportion_with_lag}
\end{figure}

\fig{fig:temporal_proportion_with_lag} shows the monthly proportion of dependencies and releases for which technical lag is strictly positive. 
We observe that the proportion of dependencies undergoing a technical lag oscillates, since 2012, between 17\% and 33\%, with a median of 24\% of all dependencies in \npm lagging behind.
The proportion of releases affected by a technical lag is higher, between 28\% and 53\% (median is 40\%), suggesting that the dependencies inducing a technical lag are spread over many different releases. 
Starting from September 2014, we start to observe a decrease in technical lag. A possible reason for this might be the increasing adoption of dependency management tools such as David DM, Gemnasium and Greenkeeper.
Trockman et al.~\cite{Trockman2018ICSE} studied the adoption of such tools by \npm packages, by analysing badges in their corresponding GitHub repositories. Among other findings, they report that dependency-manager badges signal practices that lead to fresher dependencies.

While \fig{fig:temporal_proportion_with_lag} suggests that technical lag affects a large proportion of releases, this proportion must be nuanced: not all releases are always in a situation where a technical lag is possible. 
Indeed, in order for a technical lag to affect a release, it is necessary that the latter depends on a package for which a new release is available, and that this new release does not satisfy the dependency constraint, and therefore cannot be installed automatically. 

\begin{figure}[!ht]
   \centering
   \includegraphics[width=\figsize]{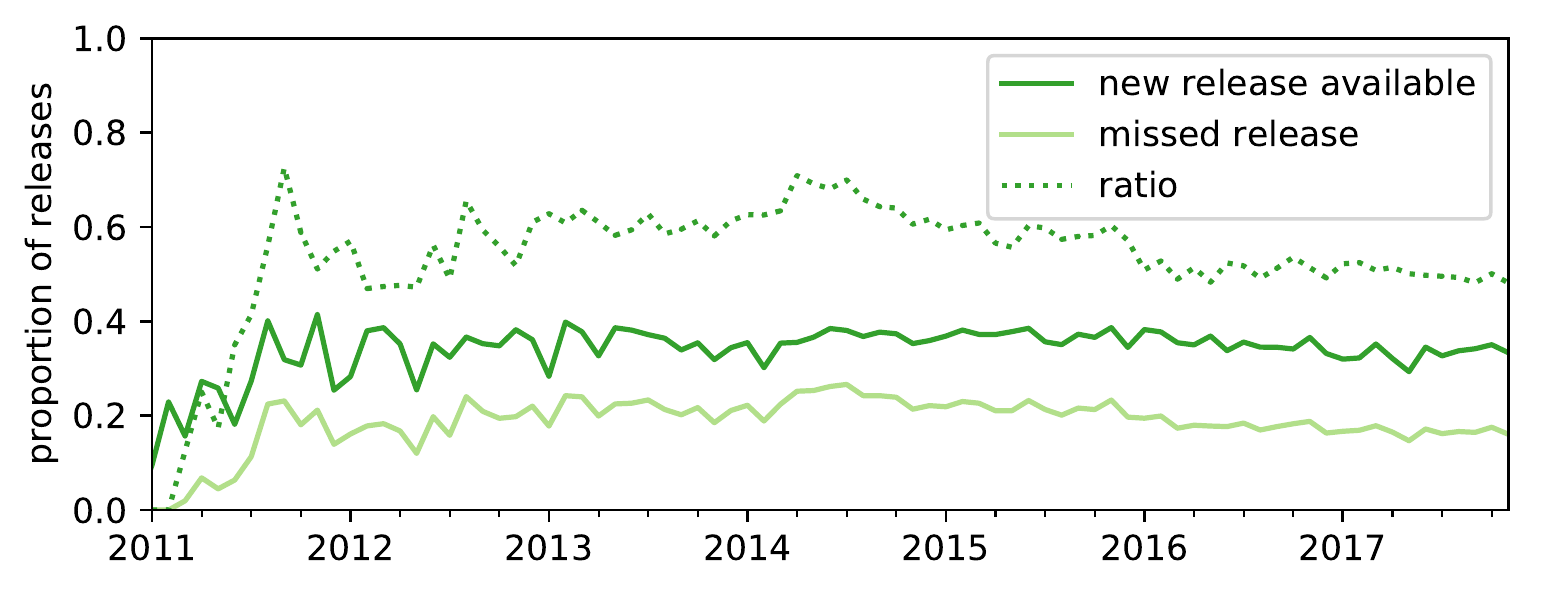}
   \caption{
     Monthly proportion of releases $r$ for which (1) a new version of a dependency target is available before $\nextt r\att{date}$, (2) a new version of a dependency target is missed before $\nextt r\att{date}$. The dotted line corresponds to the ratio of (2) over (1).
   }
   \label{fig:temporal_prop_rel_available_missed}
\end{figure}

In order to distinguish between releases that are not in a potential lag situation, those that are in a potential lag situation, and those that have effectively a technical lag, we compared, for each release $r$ and for each dependency $d\in r\att{deps}$, the sets $\available(d\att{target}, r\att{date})$ and $\available(d\att{target}, \nextt r\att{date})$. 
This allowed us to identify which are the dependency targets that were updated during the period from $r\att{date}$ to $\nextt r\att{date}$. 
Similarly, we compared $\installable(d, r\att{date})$ and $\installable(d, \nextt r\att{date})$ to identify which updates are not automatically accepted because of the dependency constraint and, therefore, induce a new technical lag. 
\fig{fig:temporal_prop_rel_available_missed} shows the monthly proportion of releases $r$ for which (1) a new version of a dependency target is available before $\nextt r\att{date}$, and (2) a new version of a dependency target is missed before $\nextt r\att{date}$. 

We observe that the proportion of releases that could have a technical lag is relatively stable over time, around 34\%. Similarly, we observe that the proportion of releases having a technical lag is stable over time, around 18\%. 
We also observe that the ratio between (1) and (2), \ie the proportion of releases having a technical lag compared to those that could have a technical lag, fluctuates from 45.7\% to 70.9\% from 2013 to 2016 (median is 60.8\%), before falling to 48.2\%-53.6\% from 2016 onwards (median is 50.8\%). 

\begin{custombox}
\textbf{Findings.} One out of four dependencies and two out of five releases suffer from technical lag. One third of all releases have at least one dependency whose target package is updated during its release life, and half of them missed this new version, inducing or increasing the technical lag. 
\end{custombox}


\section{$RQ_2$: How long is the technical lag?}
\label{sec:rq2}

$RQ_1$ focused on the extent of the technical lag in terms of affected releases and dependencies. 
A low technical lag can be explained by the time it takes for a release to make modifications to benefit from an update in its dependencies. 
With $RQ_2$ we focus on the amplitude (time delay) of that technical lag.

\begin{figure}[!ht]
   \centering
   \includegraphics[width=\figsize]{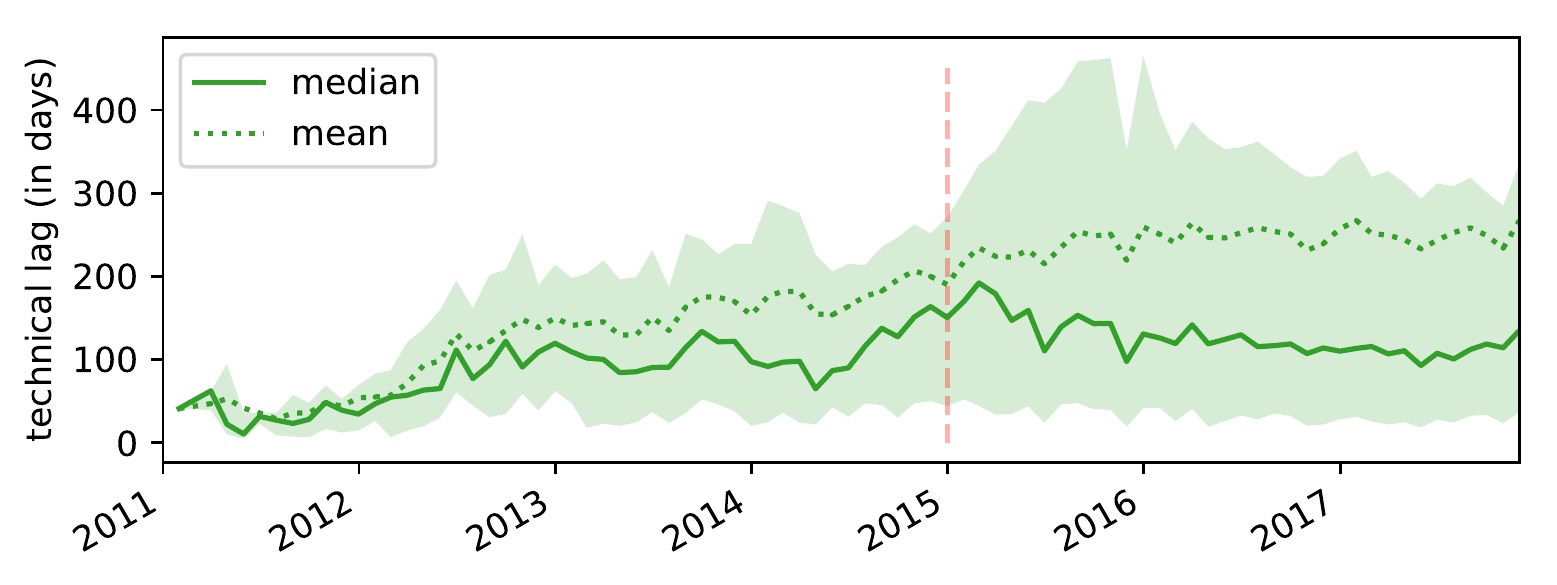}
   \caption{
     Monthly distribution of non-zero technical lag $\deltat(r,\nextt r\att{date})$. The shaded area corresponds to the interval between the 25th and 75th percentile.
   }
   \label{fig:temporal_distr_releases_at_nexttime_filtered}
\end{figure}

\fig{fig:temporal_distr_releases_at_nexttime_filtered} shows the monthly distribution of technical lag for releases with a strictly positive technical lag $\deltat(r,\nextt r\att{date})$.
We observe from \fig{fig:temporal_distr_releases_at_nexttime_filtered} that the average technical lag oscillates between 29 and 206 days until 2015.
Starting from 2015 (dashed vertical line),  the average technical lag oscillates between 215 and 267 days, with a median between 93 and 192 days, witnessing an uneven distribution of the technical lag among the releases. We also observe that since 2015,  25\% of the releases have a technical lag greater than 284 days. During the same period, only 25\% of the releases had a technical lag lower than 52 days. 

\begin{figure}[!ht]
   \centering
   \includegraphics[width=\figsize]{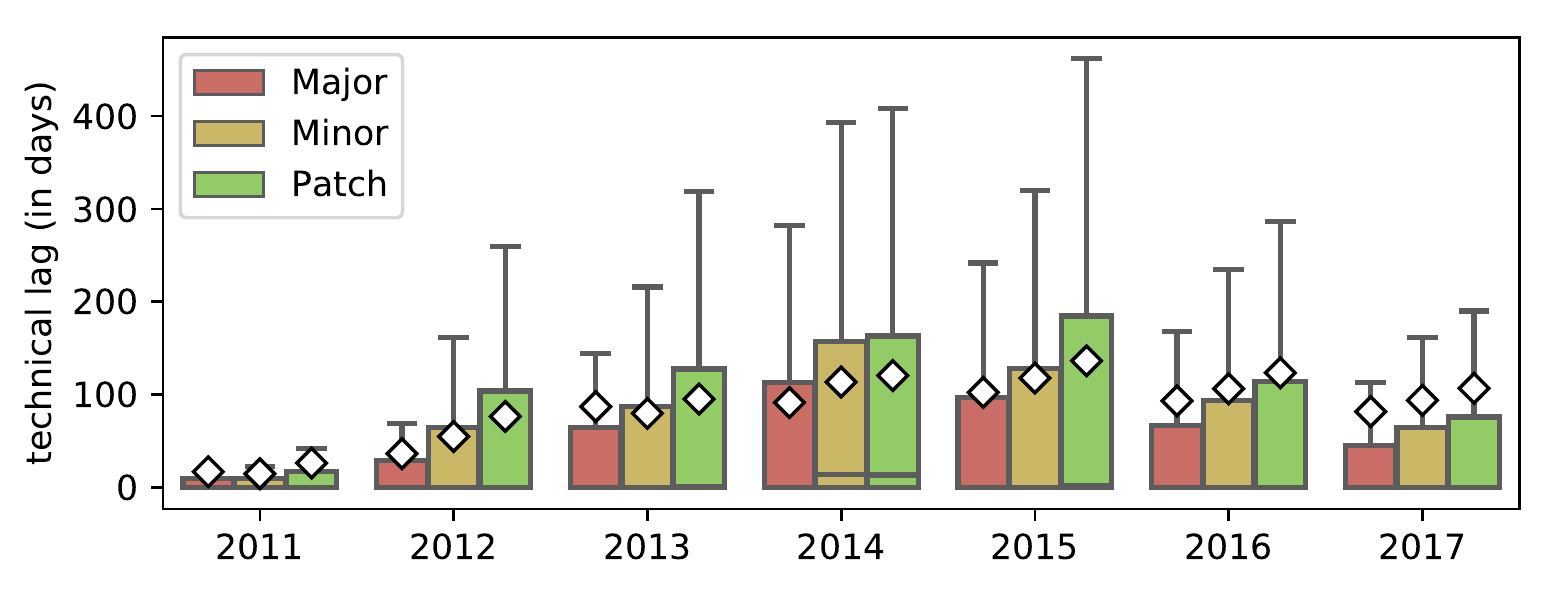}
   \caption{
     Yearly distribution of $\deltat(r,r\att{date})$ by release type.
   }
   \label{fig:temporal_distr_lag_at_next_time_per_update_type}
\end{figure}

To determine whether the release type $r_{type} \in \{major, minor, patch\}$ influences the technical lag,
\fig{fig:temporal_distr_lag_at_next_time_per_update_type} shows the technical lag distribution for $\deltat(r,r\att{date})$ per year by $r_{type}$. 
We observe that the technical lag is more important for patch releases, followed by minor releases, then major ones. 
As new updates in dependency targets can require high maintenance effort in terms of code changes, it is not surprising to observe a higher technical lag for patch and minor releases, which typically have fewer changes and are more ``lightweight updates'' than major releases.

\begin{custombox}
\textbf{Findings.} From 2015 onwards, the average technical lag for releases with technical lag is of 7 to 9 months. 25\% of the releases have a technical lag of more than 9 months, while only 25\% have a technical lag less than 52 days. 
\end{custombox}


\section{$RQ_3$: How frequently are packages updated?}
\label{sec:rq3}


Technical lag occurs when a new version of a dependency is not accepted by the dependency constraint of a release that depends on it.
Understanding the dynamics of package updates, and by extension, of dependency targets, makes it possible to better understand and identify which dependency target updates increase or decrease the technical lag of the releases that depend on them, and which releases decrease or increase the technical lag induced by their dependencies.

\begin{figure}[!ht]
   \centering
   \includegraphics[width=\figsize]{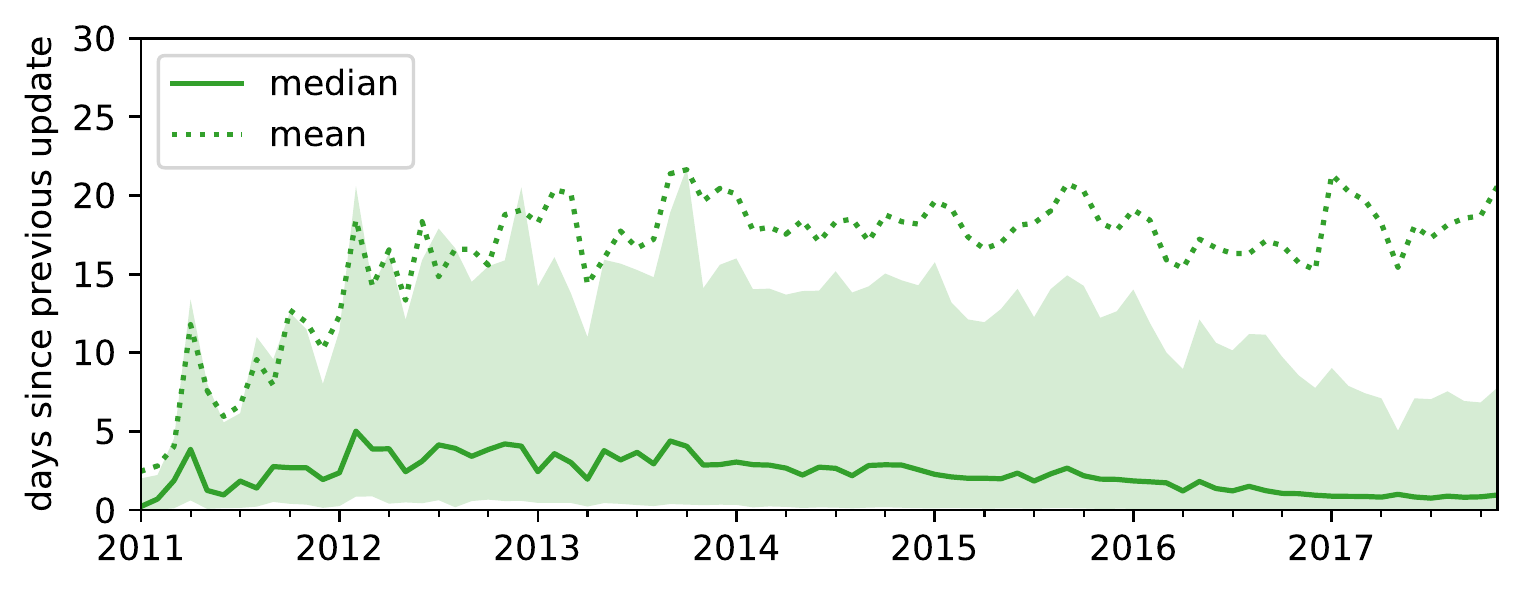}
   \caption{
     Monthly distribution of $\nextt r\att{date} - r\att{date}$.
     Shaded area corresponds to the 25th and 75th percentile.
   }
   \label{fig:temporal_distr_delta_updates}
\end{figure}

\fig{fig:temporal_distr_delta_updates} shows the monthly distribution of a release ``lifespan'', \ie the time between the date $r_{date}$ of a release $r$ and the date $\nextt r\att{date}$ of its next release.
We observe that the average time between two consecutive updates ranges from 12 to 22 days starting from 2012.
The much lower median value (1 to 5 days) suggests that the time between two consecutive updates is unevenly distributed among packages: some packages are updated very frequently (25\% within a day), and other much less frequently (25\% after 5 to 22 days).

\begin{figure}[!ht]
   \centering
   \includegraphics[width=\figsize]{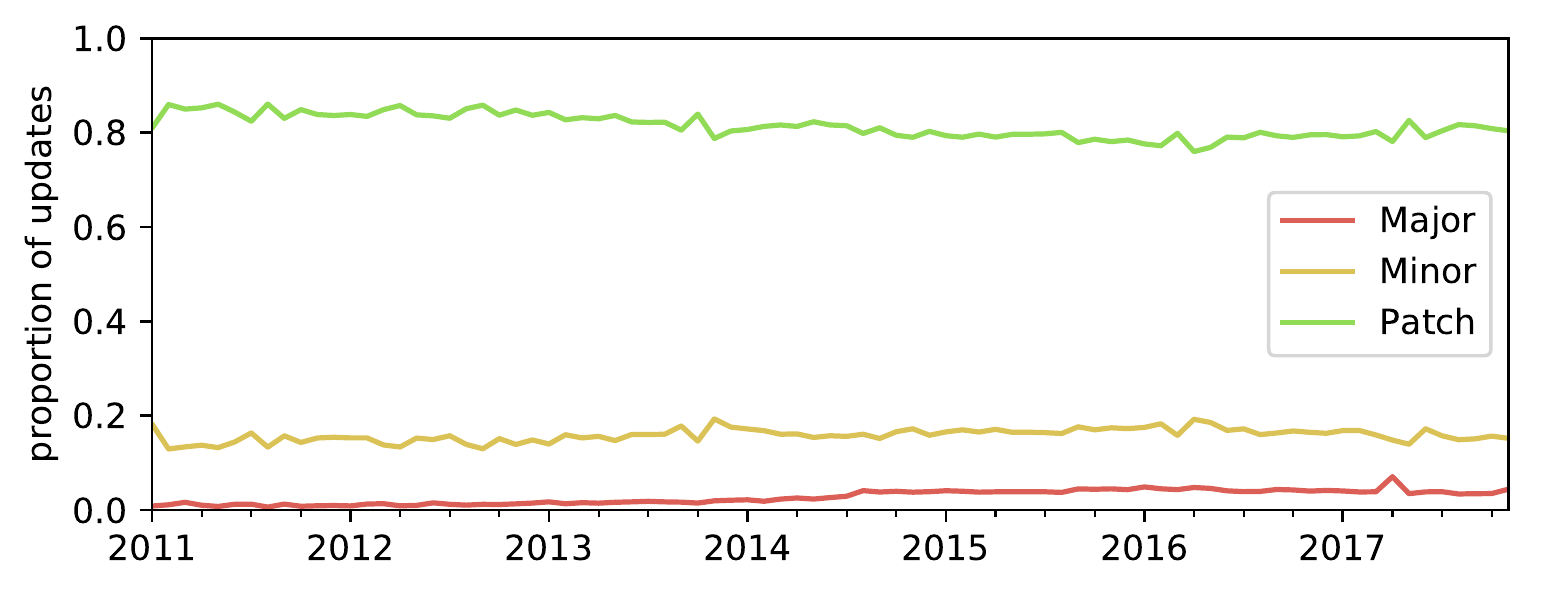}
   \caption{
     Monthly proportion of release updates per release type.
   }
   \label{fig:overview_proportion_update_type}
\end{figure}

To determine whether such a difference in update frequency can be explained by the release type, we distinguished in \fig{fig:overview_proportion_update_type} the monthly proportion of updates per release type (\ie major, minor or patch).
We observe that the vast majority (76\% to 86\%) of updates are patches, and this has been the case since the beginning of the observation period.
Minor and major updates are much less frequent, ranging from 13\% to 19\% and from 0.6\% to 7\%, respectively.

\begin{figure}[!ht]
   \centering
   \includegraphics[width=\figsize]{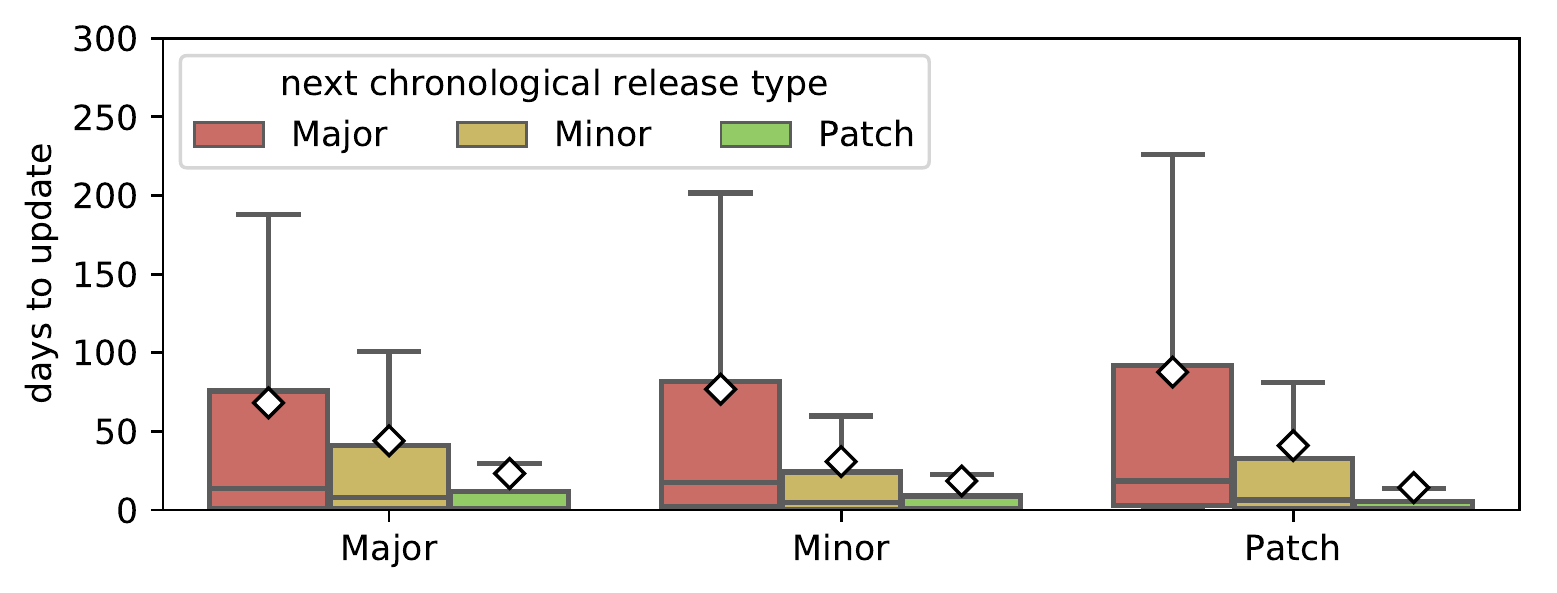}
   \caption{
     Distribution of ($\nextt r\att{date} - r\att{date}$) for all releases, grouped by release type of $r$ and $\nextt r$.
   }
   \label{fig:overview_distr_update_delay_combinations}
\end{figure}

This important distinction between the three release types suggests that we need to refine our analysis of the time between consecutive updates to take into account the release type, both for the release that is updated and for its next release (\ie the one to which a release is updated to). 
\fig{fig:overview_distr_update_delay_combinations} therefore shows the distribution of the time between two consecutive releases, by release type, and by next release type.
We observe that the time it takes to update a release essentially depends on the type of the next release.
In particular, major releases are released much longer after a previous release than minor ones. Similarly, minor releases are released much later after a previous release than patch releases.
These results are not surprising given the versioning semantics associated with each type of release. 
However, the results stress the importance of considering the release type when analysing technical lag.

\begin{custombox}
\textbf{Findings.} It takes an average of 12 to 22 days to update a release. The time required to update a release is unevenly distributed, and mainly depends on the type of the next release.
Major releases are released much later after a previous release than minor ones. Minor releases are released much later after a previous release than patches.
\end{custombox}


\section{$RQ_4$: When does technical lag increase?}
\label{sec:rq4}

While $RQ_2$ revealed that technical lag can be (very) long for many releases,
$RQ_4$ focuses on the conditions that make technical lag increase.
A release that already has technical lag at its release date will continue to see its technical lag increase during its lifespan. Therefore, it is interesting to measure the magnitude of this increase as well as its origin.

\begin{figure}[!ht]
   \centering
   \includegraphics[width=\figsize]{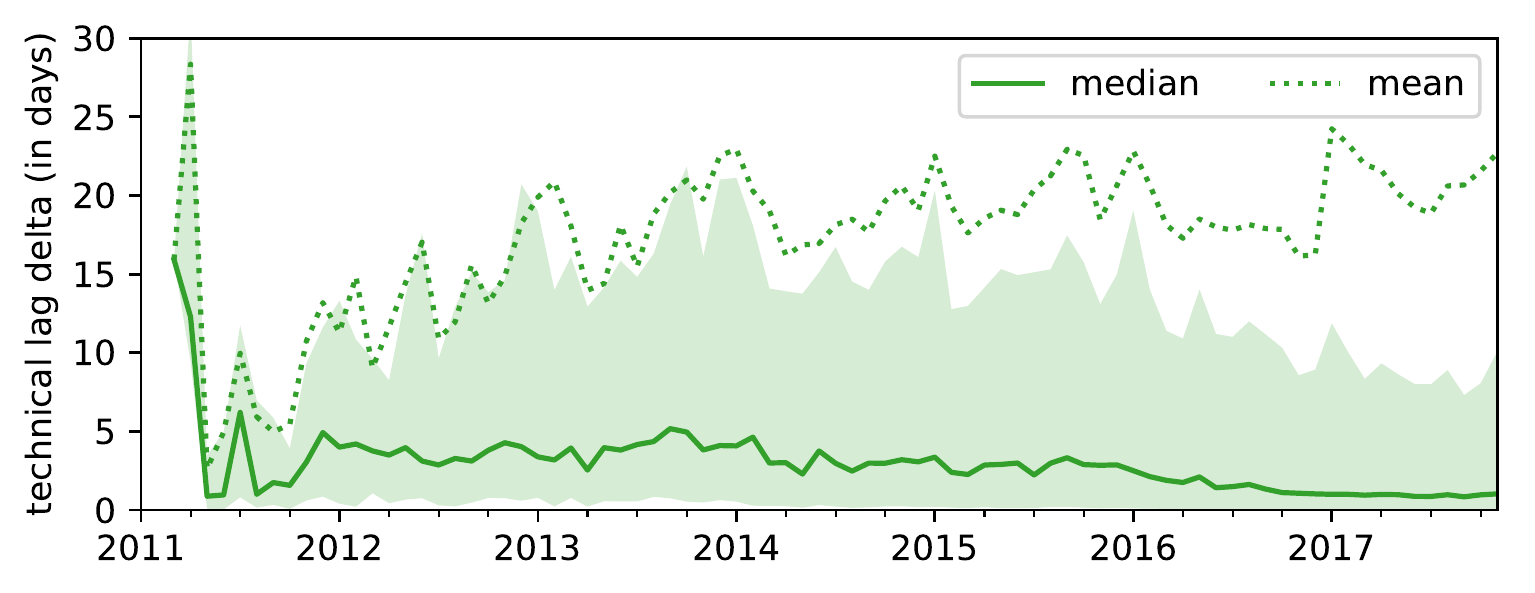}
   \caption{
     Monthly distribution of non-zero $\deltat(r, \nextt r\att{date}) - \deltat(r, r\att{date})$.
     Shaded area corresponds to the 25th and 75th percentile.
   }
   \label{fig:temporal_distr_delta_lag_releases}
\end{figure}

\fig{fig:temporal_distr_delta_lag_releases} shows the monthly distribution of the non-zero difference in technical lag $\deltat(r, \nextt r\att{date}) - \deltat(r, r\att{date})$ during the life of each release $r$.
We observe that this distribution is very similar to the one representing the time between two consecutive updates of a release (\fig{fig:temporal_distr_delta_updates}).
This is not surprising since, as \fig{fig:temporal_proportion_with_lag} already indicated, most releases $r$ having a lag at $\nextt r\att{date}$ already had a lag at $r\att{date}$.
Therefore, it is reasonable to expect that the additional lag gained between $r\att{date}$ and $\nextt r\att{date}$ corresponds to the difference between $\deltat(r, r\att{date})$ and $\deltat(r, \nextt r\att{date})$.

\begin{figure}[!ht]
   \centering
   \includegraphics[width=\figsize]{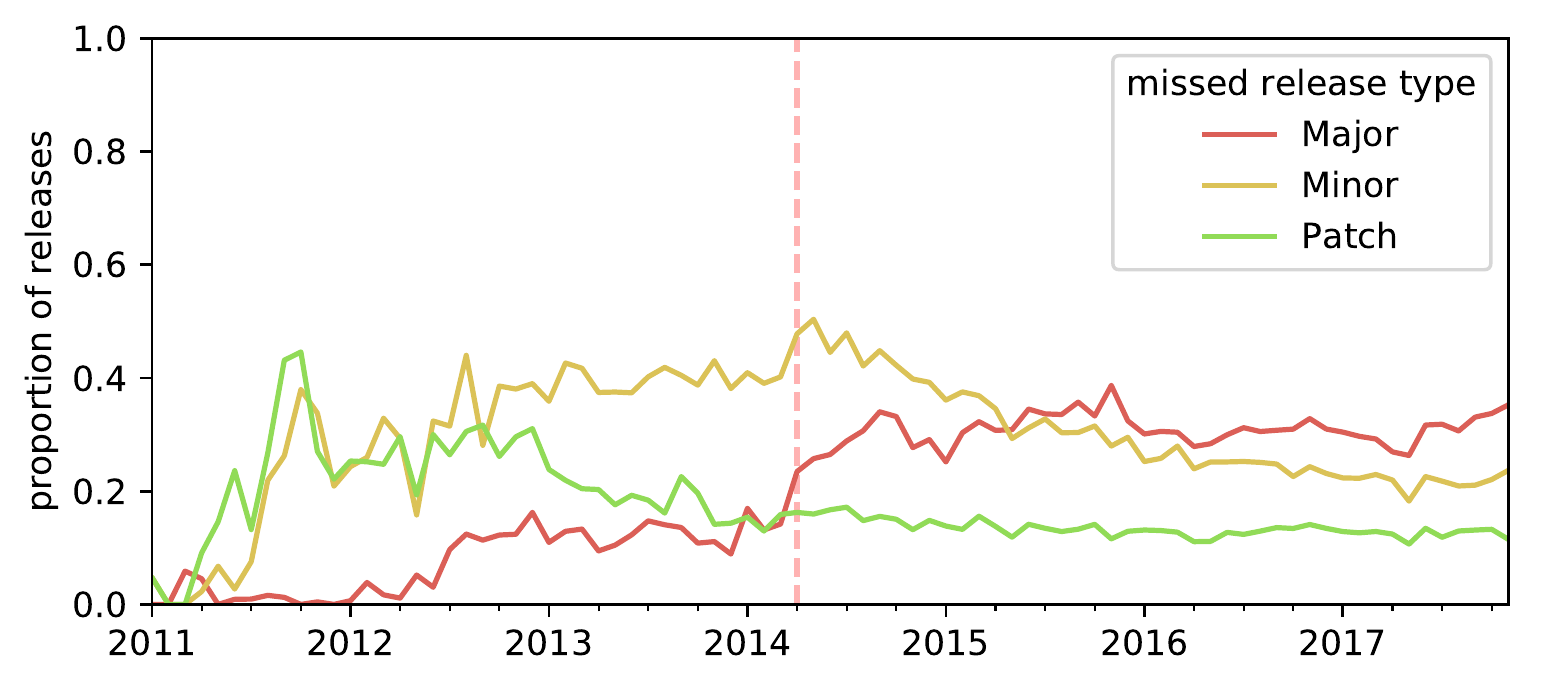}
   \caption{
     Monthly proportion of releases $r$ such that there exists $d\in r\att{deps}$ with $d\att{target}\in\missed(d, \nextt r\att{date})$. The results are grouped per release type of $d\att{target}$.
   }
   \label{fig:version_prop_missing_major_minor_patch}
\end{figure}

In order to identify which releases of a dependency target are causing this additional technical lag, we computed for each release $r$ the set of ``missed dependencies'', \ie all dependencies $d\in r\att{deps}$ such that $d\att{target}$ induces or increases the technical lag of $r$ during its lifespan, \ie  $d\att{target}\in\missed(d, \nextt r\att{date})$.
\fig{fig:version_prop_missing_major_minor_patch} shows the monthly proportion of releases $r$ in such a situation, grouped by the release type of $d\att{target}$.
We observe that technical lag is observed mostly due to minor and patch releases of dependency updates until April 2014 (vertical dashed line in \fig{fig:version_prop_missing_major_minor_patch}).
From April 2014 onwards, the proportion of major releases exceeds the one of patch releases. From mid 2015 onwards, the proportion of major releases also exceeds the one of minor releases.

These changes are probably the consequence of the introduction of a new semantic constraint \textit{caret}\footnote{https://github.com/npm/node-semver/pull/41} in \npm, its use by default in dependency constraints\footnote{\url{http://fredkschott.com/post/2014/02/npm-no-longer-defaults-to-tildes/}}, and its specific meaning for versions of the form \textsf{0.x.x}\footnote{\url{https://github.com/npm/node-semver/issues/79}}.
As a witness of the confusion induced by these changes, \npm changed the default initial version of a package from \textsf{0.1.0} to \textsf{1.0.0} since \textit{``we cannot ever hope to get everyone to believe what the correct interpretation of 0.x versions are''}\footnote{\url{https://github.com/npm/init-package-json/commit/363a17bc}}.
This has lead to an important proportion of releases that went from a \textsf{0.x.x} version scheme with only minor and patch updates, to a \textsf{1.x.x.} version. 
We computed that the proportion of releases with a \textsf{0.x.x} version dropped from 82.6\% to 66\% in 2014 alone.

The observations for \fig{fig:version_prop_missing_major_minor_patch} confirm the hypothesis that ``higher'' release types (major $>$ minor $>$ patch) require more work in their dependent packages and therefore induce additional technical lag.
It is nevertheless surprising that minor releases and (especially) patches induce technical lag since these update types are expected to be backward compatible and therefore should require nearly no effort for their adoption.

We hypothesise that the lack of automatic adoption of backward compatible changes is mainly due to the use of too strict dependency constraints, preventing patches or minor releases to be installed automatically.
Indeed,
Decan et al. observed that, in 2016, around 20\% of all \npm packages with dependency constraints specified \emph{strict} constraints, preventing new releases of a dependency to be automatically installed~\cite{Decan2017SANER}.

\begin{custombox}
\textbf{Findings.} Most of the releases with technical lag already had this lag at release date, and their technical lag continues to increase 
during their lifespan.
Most of the technical lag is due to the minor and patch releases of a dependency target. This is somehow unexpected, as minor and patch releases are supposed to be backward compatible and therefore effortless to adopt.
\end{custombox}



\section{$RQ_5$: When does technical lag decrease?}
\label{sec:rq5}

$RQ_4$ revealed that technical lag \emph{increases} continuously during the lifespan of a release.
$RQ_5$ aims to study when and how technical lag is \emph{reduced}.
To assess this, we compare for each release $r$ the technical lag at its release date $r\att{date}$ with the technical lag of its previous release $\prevt r$ at the same time, \ie $\deltat(r, r\att{date}) - \deltat(\prevt r, r\att{date})$.
A smaller technical lag implies that ``effort'' has been made to decrease the lag (\eg the dependency constraint has been adapted to accept newer releases of a dependency target).
An identical technical lag means that the dependency constraint has not been modified.
A higher technical lag means that the dependency constraint has been modified on purpose to exclude the most recently used version of a dependency target.
The latter case is expected to be extremely rare, as it corresponds to very specific situations such as the presence of a vulnerability or a bug that must be avoided as much as possible by dependent releases.

\begin{figure}[!ht]
   \centering
   \includegraphics[width=\figsize]{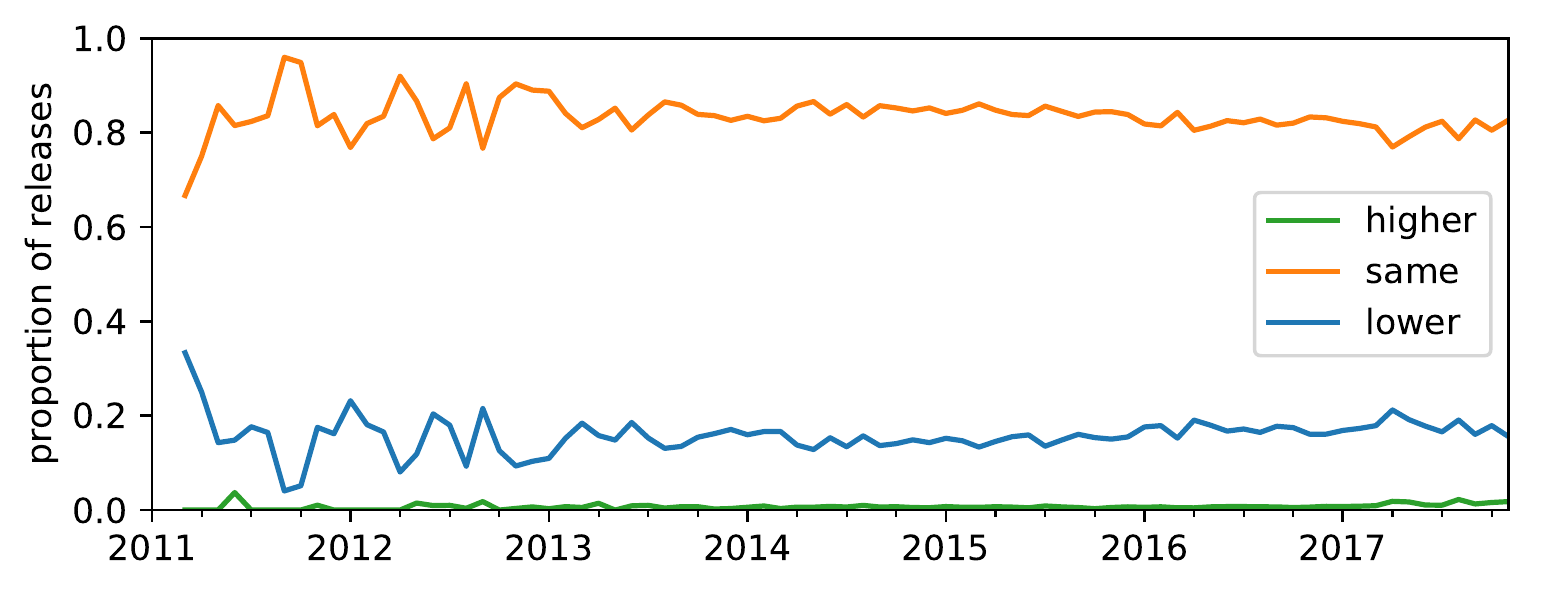}
   \caption{
     Monthly proportion of releases with a higher, similar or lower technical lag than their previous release.
   }
   \label{fig:temporal_variation_proportion}
\end{figure}

\fig{fig:temporal_variation_proportion} shows the monthly proportion of releases $r$ for which technical lag has increased, decreased or remained the same. 
We observe that these proportions remain relatively stable from 2013 onwards.
In most cases (77\% to 89\%) technical lag does not change from one release to the next.
As expected, the proportion of releases in which technical lag increases is negligible.
We observe a decrease of technical lag in only 11\% to 21\% of the cases.

\begin{figure}[!ht]
   \centering
   \includegraphics[width=\figsize]{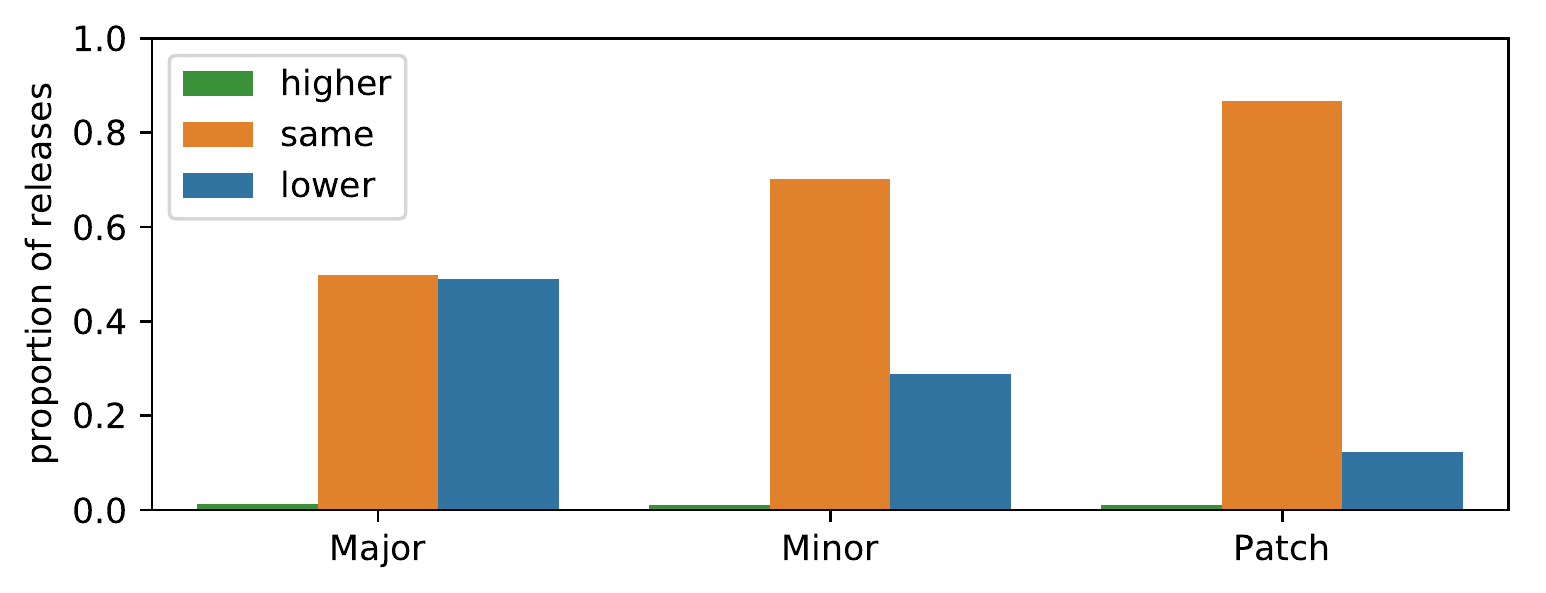}
   \caption{
     Proportion of releases (grouped per release type) with a higher, identical or lower technical lag than their previous release.
   }
   \label{fig:temporal_variation_proportion_per_type}
\end{figure}

We hypothesise that this decrease is strongly related to the release type of either $r$ or $\prevt r$.
Indeed, it seems likely that the decision to rely on a more recent release of a dependency target should be carried out as part of a larger update (\eg a major update). 
\fig{fig:temporal_variation_proportion_per_type} shows the proportion of releases of each type with a higher, identical or lower technical lag than the previous release.
It shows that a larger proportion of major releases (49\%) decrease technical lag, compared to minor (29\%) or patches (12\%).

\begin{figure}[!ht]
   \centering
   \includegraphics[width=\figsize]{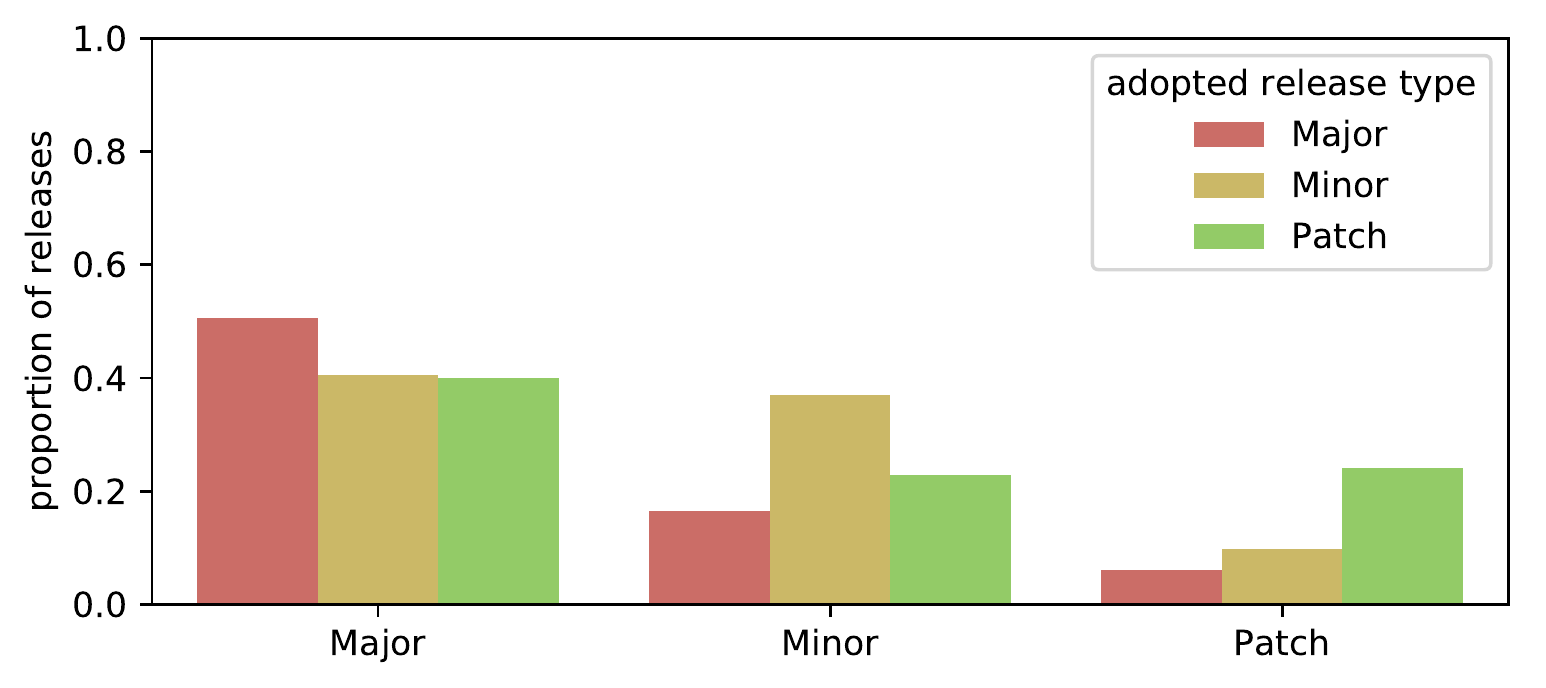}
   \caption{
     Proportion of releases (grouped by release type) adopting at least one previously missed major, minor or patch update of a dependency.
   }
   \label{fig:version_prop_adopting_per_type}
\end{figure}

\fig{fig:version_prop_adopting_per_type} shows the proportion of releases, grouped by update type, which adopted a previously missed major, minor or patch release of a dependency target.
We observe that between 40\% and 50\% of the major updates adopt a previously missed release of a dependency target.
For comparison, only around 25\% of the missed releases are adopted during minor updates, and only 13\% during patch updates.

We expected major releases that were missed to be adopted during major updates, as they usually contain backward incompatible changes and require additional efforts to be adopted. However, we did not expect such low proportions of adopted minor and patch releases.
Only 23\% (resp. 24\%) of previously missed patch releases are adopted during a minor (resp. patch) update, versus 40\% during major updates.
This is surprising, as they are expected to be backward compatible and thus effortless to adopt, especially patches that should be adopted quickly as they are supposed to fix bugs and security issues.

\begin{custombox}
\textbf{Findings.} Most of the technical lag is reduced during major updates (and, to a lesser extent, during minor updates). Major releases of a dependency are typically adopted during a major update, and minor releases during a major or a minor update. Less than one third of all backward compatible releases are adopted during minor or patch updates.
\end{custombox}


\section{$RQ_6$: How could technical lag be reduced by proper use of semantic versioning?}
\label{sec:rq6}

Let us reconsider \fig{fig:version_prop_adopting_per_type} in the light of semantic versioning.
It revealed that a relatively low proportion of releases adopt minor or patch updates of their dependencies.

Since patch releases are typically used for fixing bugs or security issues, it is not surprising that they do not represent an ``ideal time`` to update dependencies and to adopt new releases, even if these are expected to be backward compatible.
Minor releases, however, could be reasonably expected to reduce their technical lag by making use of more recent minor or patch updates of their dependencies.
In fact, using tilde $\sim$ or caret $\string^$ in the dependency constraint would enable this.
For example, ``$\sim1.2.3$" permits releases up to the next \emph{minor} release (excluded).
Similarly, ``$\string^1.2.3$" permits releases up to the next \emph{major} release (excluded).
If, in contrast, strict or maximal constraints (\eg ``$1.2.3$" or ``$<= 1.2.3$'') would be used, no recent updates of the dependency will be accepted.

\begin{figure}[!ht]
   \centering
\includegraphics[width=\figsize]{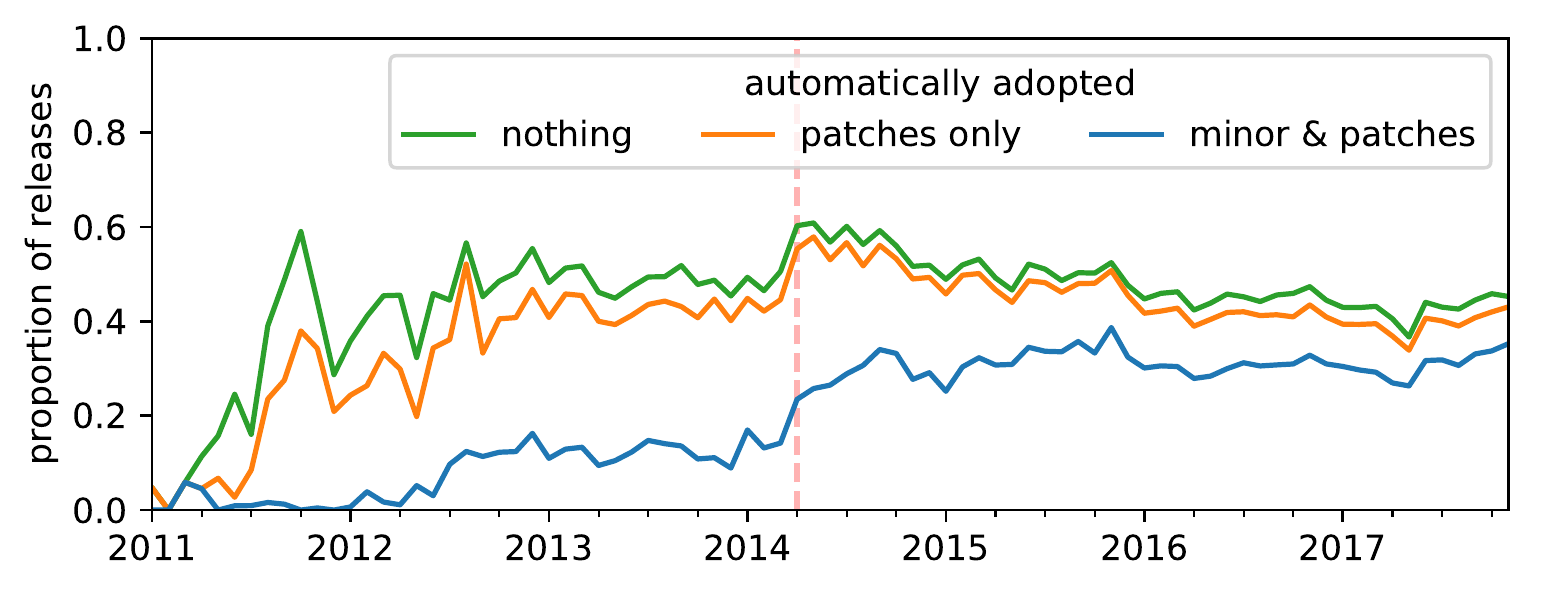}
\caption{Monthly proportion of releases $r$ having a technical lag $\deltat(r, \nextt r\att{date}) >0$ under the ``what if'' analysis.}
   \label{fig:whatif}
\end{figure}

In order to assess to which extent a proper use of semantic versioning could help to reduce the effect of technical lag, we carried out a ``what if'' analysis. \fig{fig:whatif} shows what would happen if dependency constraints would be ``loosened'' to allow for either patches, or both patches and minor releases to be accepted automatically.
While we observe that the automatic adoption of patches would not change much, allowing (backward compatible) minor releases and patches would reduce the proportion of releases suffering from technical lag.
Indeed, from April 2014 onwards (vertical dashed line in \fig{fig:whatif}), allowing backward compatible releases lowers the proportion of releases suffering from technical lag by an average of 17.6\%.

\begin{custombox}
\textbf{Findings.} If dependency constraints would rely on semantic versioning rules 
that enable automatic updates of backward compatible changes, the proportion of releases suffering from technical lag could be reduced by 17.7\%.
\end{custombox}


\section{Discussion}
\label{sec:discussion}

Our empirical results revealed that many package releases exhibit technical lag. Package maintainers may impose too strict dependency constraints that refrain from updating their package dependencies to more recent versions, possibly because they are concerned with the extra effort or risk it would entail.
However, assuming that package developers respect the semantic versioning policy, there is nothing that should prevent them from benefiting from backward compatible updates provided through minor or patch updates.
In fact, enabling such a more flexible update policy has shown to be advantageous.
For example, Cox \etal~\cite{Cox2015} observed that up-to-date systems are four times less likely to suffer from security issues and backward incompatibilities than systems that are up-to-date.

\begin{figure}[!t]
   \centering
   \includegraphics[width=\figsize]{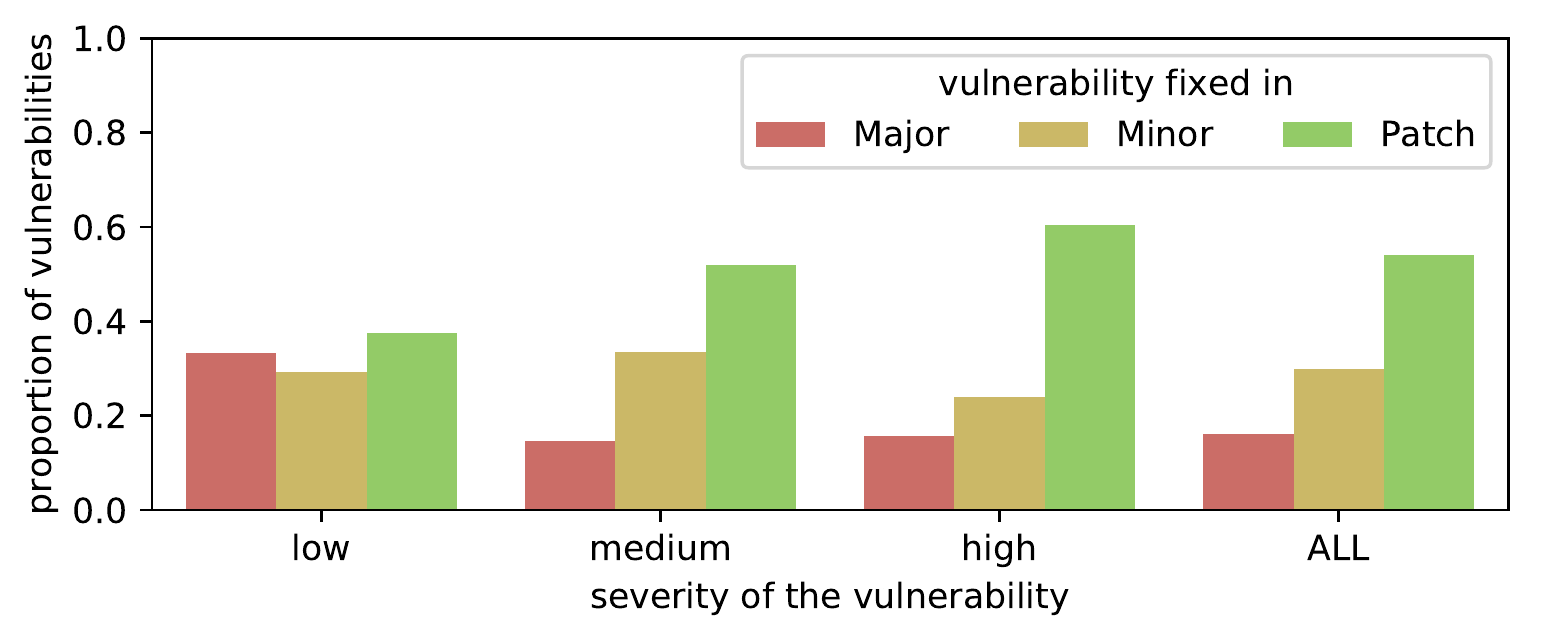}
   \caption{Proportion of vulnerabilities fixed in patch, minor or major releases.}
   \label{fig:vuln_update_fix}
\end{figure}

In earlier work we studied the evolution of vulnerabilities in \npm~\cite{Decan2018MSR}. Based on these results, we assessed to which extent automatic updates to patches or minor releases would allow dependent releases to benefit from fixes to vulnerabilities. The results are shown in \fig{fig:vuln_update_fix}.
Most vulnerabilities are indeed fixed during a patch (54\%) or a minor release (30\%), while only 16\% are fixed during a major release.

\begin{custombox}
\textbf{Actionable result.} Package maintainers should use semantic versioning 
to benefit from automatic backward compatible updates of patches and minor releases.
This would reduce the risk of suffering from vulnerabilities as most of them are fixed during minor and patch updates.
\end{custombox}

In response to $RQ_1$, we observed a decrease in the monthly proportion of package releases and dependencies suffering from technical lag.
Dependency management tools such as David, Gemnasium and Greenkeeper help maintainers to keep their dependencies up-to-date, by 
suggesting updates when new releases of dependent packages become available.
By analysing badges in \github repositories, Trockman et al.~\cite{Trockman2018ICSE} studied the adoption by \npm packages of dependency management tools.
Based on interviews with developers they concluded that dependency-management badges signal practices that lead to fresher dependencies, indicating attention to updates and security patches.

\begin{custombox}
\textbf{Actionable result.} Dependency management tools help package maintainers to reduce the technical lag of packages.
\end{custombox}

While it is useful for package maintainers to know if their dependencies are up to date, the \npm community as a whole can also benefit from having an ecosystem-wide view of the impact of technical lag. Indeed, our analyses have revealed that some changes in the \npm policy may lead to important changes in technical lag over time. This was for example the case when \npm introduced changes to the semantic constraint caret, causing an important shift of packages from the 0.x.x to the 1.x.x version scheme (see \sect{sec:rq4}).

Another example is the presence of unmaintained packages (\eg because its maintainers are no longer available). Ecosystem managers should reduce such cases to a minimum. Indeed, if unmaintained packages depend on other packages, their technical lag will continue to increase over time, since there is nobody that will update the package if new releases of dependent packages appear. As a consequence, these packages will incur an increased risk of bugs and security issues, as there is nobody that will monitor and fix these problems. 

\begin{custombox}
\textbf{Actionable result.} Ecosystem managers should adopt an ecosystem-wide view of technical lag, in order to monitor its temporal evolution, as well as the impact of changes in the package distribution policy or tooling.
\end{custombox}

The \npmsio service (\url{https://www.npms.io}) continuously computes a  \emph{popularity}, \emph{quality} and \emph{maintenance} score for each \npm package, using information from a variety of sources, such as \github, the David dependency manager, and the Node Security Platform. 
The number of outdated dependencies is used as one of the factors to compute these scores, but this is not sufficient to provide a historic view of the dependency freshness of each package.
It would therefore be useful to include the technical lag of package releases as a factor of package maintainability, as it indicates the ability of the package to remain up to date with respect to its dependencies.

\begin{custombox}
\textbf{Actionable result.} Dependency monitoring tools and even the \npm package manager should incorporate information about the technical lag of installed packages.
\end{custombox}

As observed in \fig{fig:version_prop_missing_major_minor_patch}, 
since mid 2015 major releases are the most important cause of technical lag. This is not surprising, since backward incompatible changes are only expected to take place in major releases (assuming semantic versioning policy is respected).
Since dependent packages are less likely to upgrade to major releases of their dependencies,  
package maintainers should take this into account. For example, if an important bug or security vulnerability is detected that is also relevant for earlier releases, the maintainer should strive to backport the fixes to those previous releases that are known to be required by other packages. This will reduce the risk and technical lag of those dependent packages that did not upgrade yet to the most recent major release.

\begin{custombox}
\textbf{Actionable result.} When providing a new major release, maintainers should strive to support dependent packages to update to such a release as easily as possible. They should also strive to backport important fixes to earlier releases.
\end{custombox}

\section{Threats to validity}
\label{sec:threatsthreat}


The accuracy of our results relies on the correctness of the package dependency metadata extracted from \textsf{libraries.io}.
However, this metadata was manually checked for correctness in previous work~\cite{DecanEMSE2018}.

It may also be the case that some packages (and their history) have been removed from \npm before the extraction date, in which case they are not considered during our analysis.
This is now prohibited by \npm since April 2016.
\footnote{\url{http://blog.npmjs.org/post/141905368000/changes-to-npms-unpublish}}
This threat is unlikely to affect our results because of the huge number of packages we considered, and because we filtered out all dependencies pointing to non-existing package releases.

We restricted the dependencies of the \npm package dependency network to those required to install and execute the package (\ie runtime dependencies).
Dependencies that are only required to develop or test a package were excluded from our analyses because not every package declares a complete and reliable list of development or test dependencies, and because these dependencies are unlikely to affect the production environment where technical lag could matter.

Another threat relates to the dates we considered when computing the technical lag of a release.
For each package we measured the technical lag at the date of each release (\ie the initial technical lag) and at the date of the next release, assuming that this represents the expected ``end-of-life'' of the release.
In practice, a user can install or depend explicitly on a release that is not the latest available one. In that case, the technical lag could be higher than the one we computed for that release.
Since this potentially higher technical lag is a consequence of an explicit choice made by the user (and not by the maintainer of the package), we consider that it should not be taken into account in our analyses.
The values we reported are therefore more representative of the inherent technical lag of a package, and can be considered as a lower bound of the technical lag that can be observed in practice.

A final threat relates to the generalisability of our findings. The approach could be replicated for dependency networks of other package distributions, but the findings may be quite different from those obtained for \npm, due to the fact that each package distribution and community has different policies, practices and culture \cite{Bogart2016,Decan2017SANER}.


\section{Related Work}
\label{sec:related}

Many researchers have studied the evolution characteristics of package dependency networks.
Wittern \etal~\cite{Wittern2016} analysed different evolution characteristics of \npm packages, such as their dependencies, update frequency, popularity and versioning policy. They observed that maintainers use different versioning conventions for their packages, that are not always compatible with the recommended semantic versioning policy. This practice resulted in different version adoption ratios.
%
Decan \etal \cite{Decan2017SANER} compared the topology of \npm with the one of the \cran and \rubygems  package dependency networks.
They studied the use of dependency constraints and found that, while strict dependency constraints increase the risk of missing important updates, they also prevent backwards incompatibility issues.
%
%
Raemaekers \etal \cite{Raemaekers2014} investigated the use of semantic versioning in 22k Java libraries in \maven over a seven-year time period. They found that breaking changes appear in one third of all releases, including minor releases and patches. Somehow surprisingly, they observed that breaking changes have little influence on the actual delay between the availability of a release and its adoption by dependent packages.
Bogart \etal \cite{Bogart2016} conducted a qualitative comparison of \npm, \cran and \eclipse, to understand the impact of community values, tools and policies on breaking changes. By interviewing developers, they found that there are two main types of mitigation strategies to reduce the exposure to changes in dependencies: limiting the number of dependencies, or depending only on ``trusted'' packages.

Many researchers have studied the phenomenon of outdated dependencies.
McDonnell \etal \cite{McDonnell2013-ICSM} studied the evolution of the Android API, and the adoption of API updates by client applications. Among other findings, they observed that about 28\% of API references in client applications are outdated, with a median time lag of 16 months. 22\% of these outdated API usages eventually upgrade to newer API versions, at a much slower rate than the average API release interval.
Kula et al.~\cite{Kula2016SANER} analyzed over 6k Java libraries in \maven to investigate the latency in adopting the latest release of dependency targets.
They showed that maintainers are more likely adopt the latest release for newly introduced dependencies, but less likely to adopt them at the beginning of their projects.
In a follow-up study~\cite{Kula2018EMSE}, they investigated 4.6k \github projects with 2.7k library dependencies and found that more than 80\% of the studied systems have outdated dependencies.

Cox et al.~\cite{Cox2015} proposed different metrics to quantify a software system's \emph{dependency freshness}. They measured the version sequence number, release date and number delta to capture four criteria, namely technology independence, ease of implementation, simplicity to understand, and enable root-cause analysis.
They assessed the usefulness of these metrics through interviews with five technical consultants at SIG and showed that their objective metrics are in agreement with the subjective perception of dependency freshness.
%
%

Gonzalez-Barahona et al.~\cite{Barahona2017OSS} introduced the concept of technical lag to measure how outdated a system is with respect to its dependencies.
They defined technical lag in terms of a lag function and lag aggregation function for packages.
Zerouali et al.~\cite{Zerouali2018} measured such technical lag for \npm package dependencies in terms of time and of number of missed versions. They observed that a large number of dependencies in \npm have a technical lag of several months. However, our analysis differs from theirs in many ways. First of all, 
we explicitly excluded development dependencies as they tend to lead to an overapproximation of technical lag. We also computed technical lag not only for dependencies, but also for releases, as an aggregation of the technical lag of its dependencies.
The way in which they computed technical lag, by comparing the latest installable release with the latest available one, is not consistent with how the \npm package manager works, leading to inaccurate results when multiple ``branches'' of a package are maintained in parallel, a common phenomenon for popular \npm packages. Instead, one should consider not the \emph{latest} release, but the release corresponding to the \emph{highest version}.
Another difference is that we computed technical lag at the release dates of both the current \emph{and} the next release of each package, allowing us to provide insights about the change in technical lag during a release lifespan, to assess the variation in technical lag along subsequent releases, and to identify when technical lag is reduced by a new release.

The phenomenon of technical lag and outdated dependencies has been shown to increase the risk of security vulnerabilities.
Cox et al.~\cite{Cox2015} analyzed 75 Java systems in \maven, and split them into four risk categories based on their \emph{dependency freshness}. They compared the systems in each category w.r.t. reported vulnerabilities and found that systems with a low dependency freshness are more than four times as likely to contain security issues in these dependencies.
Derr et al.~\cite{Derr2017CCS} conducted a survey with more than 200 app developers in the Android ecosystem to
investigate the use of outdated libraries, and reported that 
almost 98\% of 17k actively used library versions with a known security vulnerability could be easily fixed by updating the library to a fixed version.
Decan et al.~\cite{Decan2018MSR} empirically studied nearly 400 security reports for 269 \npm packages.
They found that these vulnerabilities affect more than 72k other packages due to dependencies.
They also reported that more than 40\% of the releases depending on a vulnerable package do not automatically benefit from the security fixes because of too restrictive dependency constraints.

A large body of research focused on the evolution of library APIs.
Wu \etal \cite{Wu2016} studied the API evolution in Apache and Eclipse. Raemaekers \etal \cite{Raemaekers2014} analysed breaking changes and deprecated methods in Maven packages in presence of semantic versioning. Robbes \etal \cite{Robbes2012-smalltalkecos} empirically studied the ripple effect of deprecated APIs in the Smalltalk ecosystem. 
Hora \etal \cite{Hora2018} explored the evolution of the Pharo ecosystem. This line of research differs from our own by its level of granularity: by analysing fine-grained API changes (e.g., at method level), it becomes possible to assess the effort and impact required to upgrade client applications to newer API versions.
Carrying out such fine-grained static analyses is not feasible at the level of the entire \npm ecosystem, because of the highly dynamic nature of the JavaScript language. While partial solutions and heuristics have been proposed (e.g. through combinations of pointer analysis and use analysis \cite{Madsen2013PSA}), there are still many remaining open problems and challenges that need to be overcome \cite{Sun2017AJP} to make it scale.

\section{Future Work}
\label{sec:future}

We plan to replicate our study on package dependency networks of different package distributions. This will allow us to compare the extent of technical lag across different ecosystems, in order to gain a better understanding of how the specific policies, culture and tools affect the presence of technical lag and its evolution over time.
Inspired by \cite{Decan2017SANER}, who studied the use of dependency constraints in three different package distributions (\npm, \cran and \rubygems), we aim to explore to which extent the specific (semantic) versioning policy and the use of dependency constraints influences the presence of technical lag.


Inspired by \cite{Cox2015,Derr2017CCS,Decan2018MSR}, we aim to empirically study the negative aspects of technical lag, such as the increased risk of suffering from security vulnerabilities and bugs.
%
We also aim to understand the main causes of technical lag in packages, by quantitatively studying the relation between technical lag and a wide range of socio-technical package characteristics such as their age, code size, developer community, usage popularity, and update frequency.
This quantitative analysis will be complemented by a qualitative survey, targeting maintainers of \emph{relevant} packages (\eg packages that have a high technical lag, that induce technical lag on their dependents, that quickly reacting to dependency updates, etc.). With the results of this survey, we aim to identify and understand the reasons that lead developers to manage dependency updates in a specific way (\eg why do some package maintainters update dependencies mainly during major updates, why don't they use dependency constraints that automatically allow for backward compatible releases, etc.).
Such a study would complement the interviews carried out by Cox et al.~\cite{Cox2015}, who assessed dependency freshness awareness by developers and validated the relevance and utility of several dependency freshness related metrics through interviews with five technical consultants.



\section{Conclusion}
\label{sec:conclusion}

Package distributions, like \npm for JavaScript, are composed of huge interdependent collections of reusable software packages. These packages can suffer from technical lag if they depend on outdated packages, for example because the imposed dependency constraints prevent the package from installing a more recent version of its dependencies. Technical lag reflects the duration of time during which a package remains out of date with respect to a dependent package.

Based on the \textsf{libraries.io} dataset, we carried a longitudinal empirical study of such technical lag for 120k packages in the \npm package dependency network, over an eight-year time period.
We analysed how many packages exhibit technical lag over time, how widespread this lag is over the entire \npm package distribution, and which types of package releases are more subject to technical lag depending on their ``semantic'' version (major, minor or patch).
We also explored when, and for which types of releases (major, minor or patch), technical lag tends to increase over time, and which types of release updates (major, minor or patch) tend to reduce technical lag.

We observed that package releases suffer from technical lag if they do not benefit from the latest updates of a dependency target. It takes on average less than three weeks for a package to be updated, but this time is unevenly distributed and also depends on the type of the update (major, minor or patch).
Large proportions of dependencies and releases of \npm packages suffer from technical lag because of these updates in dependency targets.
Since 2015, the technical lag of package releases ranges between 7 to 9 months.
Technical lag is mainly reduced during major updates, even if it is mainly due to minor and patch releases that are supposed to be backward compatible and thus effortless to adopt.
A proper use of semantic versioning would clearly help to further decrease the effect of technical lag. A ``what if'' analysis revealed that the proportion of releases suffering from technical lag could be reduced by nearly 18\% if backward compatible updates of a dependency target would be automatically adopted.

\section{Acknowledgements}
This research was carried out in the context of FRQ-FNRS collaborative research project R.60.04.18.F ``SECOHealth'', FNRS Research Credit J.0023.16 ``Analysis of Software Project Survival'' and Excellence of Science project 30446992 \textsf{SECO-Assist} financed by FWO - Vlaanderen and F.R.S.-FNRS.

\bibliographystyle{IEEEtran}
\bibliography{biblio}

\providecommand{\noopsort}[1]{}
\begin{thebibliography}{10}
\providecommand{\url}[1]{#1}
\csname url@samestyle\endcsname
\providecommand{\newblock}{\relax}
\providecommand{\bibinfo}[2]{#2}
\providecommand{\BIBentrySTDinterwordspacing}{\spaceskip=0pt\relax}
\providecommand{\BIBentryALTinterwordstretchfactor}{4}
\providecommand{\BIBentryALTinterwordspacing}{\spaceskip=\fontdimen2\font plus
\BIBentryALTinterwordstretchfactor\fontdimen3\font minus
  \fontdimen4\font\relax}
\providecommand{\BIBforeignlanguage}[2]{{%
\expandafter\ifx\csname l@#1\endcsname\relax
\typeout{** WARNING: IEEEtran.bst: No hyphenation pattern has been}%
\typeout{** loaded for the language `#1'. Using the pattern for}%
\typeout{** the default language instead.}%
\else
\language=\csname l@#1\endcsname
\fi
#2}}
\providecommand{\BIBdecl}{\relax}
\BIBdecl

\bibitem{Barahona2017OSS}
J.~M. Gonzalez-Barahona, P.~Sherwood, G.~Robles, and D.~Izquierdo, ``Technical
  lag in software compilations: Measuring how outdated a software deployment
  is,'' in \emph{IFIP International Conf. on Open Source Systems}, 2017, pp.
  182--192.

\bibitem{Zerouali2018}
A.~Zerouali, E.~Constantinou, T.~Mens, G.~Robles, and J.~Gonzalez-Barahona,
  ``{An Empirical Analysis of Technical Lag in npm Package Dependencies},'' in
  \emph{Int'l Conf. Software Reuse ({ICSR})}, 2018.

\bibitem{Kula2018EMSE}
R.~G. Kula, D.~M. German, A.~Ouni, T.~Ishio, and K.~Inoue, ``Do developers
  update their library dependencies?'' \emph{Empirical Software Engineering},
  vol.~23, no.~1, pp. 384--417, Feb. 2018.

\bibitem{Bogart2016}
C.~Bogart, C.~K\"astner, J.~Herbsleb, and F.~Thung, ``How to break an {API}:
  Cost negotiation and community values in three software ecosystems,'' in
  \emph{Int'l Symp. Foundations of Software Engineering}, 2016.

\bibitem{DecanEMSE2018}
A.~Decan, T.~Mens, and P.~Grosjean, ``An empirical comparison of dependency
  network evolution in seven software packaging ecosystems,'' \emph{Empirical
  Software Engineering}, Feb 2018.

\bibitem{andrew_nesbitt_2017_808273}
\BIBentryALTinterwordspacing
A.~Nesbitt and B.~Nickolls, ``{Libraries.io} open source repository and
  dependency metadata,'' Jun. 2017. [Online]. Available:
  \url{https://doi.org/10.5281/zenodo.808273}
\BIBentrySTDinterwordspacing

\bibitem{Trockman2018ICSE}
A.~Trockman, S.~Zhou, C.~K\"{a}stner, and B.~Vasilescu, ``Adding sparkle to
  social coding: An empirical study of repository badges in the npm
  ecosystem,'' in \emph{International Conference on Software Engineering
  (ICSE)}, 2018.

\bibitem{Decan2017SANER}
A.~Decan, T.~Mens, and M.~Claes, ``An empirical comparison of dependency issues
  in {OSS} packaging ecosystems,'' in \emph{Int'l Conf. Software Analysis,
  Evolution, and Reengineering}, 2017, pp. 2--12.

\bibitem{Cox2015}
J.~Cox, E.~Bouwers, M.~van Eekelen, and J.~Visser, ``Measuring dependency
  freshness in software systems,'' in \emph{Int'l Conf. Software
  Engineering}.\hskip 1em plus 0.5em minus 0.4em\relax IEEE Press, 2015, pp.
  109--118.

\bibitem{Decan2018MSR}
A.~Decan, T.~Mens, and E.~Constantinou, ``On the impact of security
  vulnerabilities in the npm package dependency network,'' in
  \emph{International Conference on Mining Software Repositories}, May 2018.

\bibitem{Wittern2016}
E.~Wittern, P.~Suter, and S.~Rajagopalan, ``A look at the dynamics of the
  {JavaScript} package ecosystem,'' in \emph{Int'l Conf. Mining Software
  Repositories}.\hskip 1em plus 0.5em minus 0.4em\relax ACM, 2016, pp.
  351--361.

\bibitem{Raemaekers2014}
S.~Raemaekers, A.~{van Deursen}, and J.~Visser, ``Semantic versioning versus
  breaking changes: A study of the {Maven} repository,'' in \emph{Working Conf.
  Source Code Analysis and Manipulation}, Sept 2014, pp. 215--224.

\bibitem{McDonnell2013-ICSM}
T.~McDonnell, B.~Ray, and M.~Kim, ``An empirical study of {API} stability and
  adoption in the {Android} ecosystem,'' in \emph{International Conference on
  Software Maintenance}, 2013, pp. 70--79.

\bibitem{Kula2016SANER}
R.~G. Kula, D.~M. German, T.~Ishio, and K.~Inoue, ``Trusting a library: A study
  of the latency to adopt the latest {Maven} release,'' in \emph{Int'l Conf. on
  Software Analysis, Evolution, and Reengineering}, March 2015, pp. 520--524.

\bibitem{Derr2017CCS}
E.~Derr, S.~Bugiel, S.~Fahl, Y.~Acar, and M.~Backes, ``Keep me updated: An
  empirical study of third-party library updatability on {Android},'' in
  \emph{ACM Conf. on Computer and Communications Security}, October 2017.

\bibitem{Wu2016}
W.~Wu, F.~Khomh, B.~Adams, Y.-G. Gu{\'e}h{\'e}neuc, and G.~Antoniol, ``An
  exploratory study of {API} changes and usages based on {Apache} and {Eclipse}
  ecosystems,'' \emph{Empirical Software Engineering}, vol.~21, no.~6, pp.
  2366--2412, Dec 2016.

\bibitem{Robbes2012-smalltalkecos}
R.~Robbes, M.~Lungu, and D.~R\"{o}thlisberger, ``How do developers react to
  {API} deprecation? {The} case of a {Smalltalk} ecosystem,'' in \emph{Int'l
  Symp. Foundations of Software Engineering}.\hskip 1em plus 0.5em minus
  0.4em\relax { ACM }, 2012.

\bibitem{Hora2018}
A.~Hora, R.~Robbes, M.~T. Valente, N.~Anquetil, A.~Etien, and S.~Ducasse, ``How
  do developers react to {API} evolution? {A} large-scale empirical study,''
  \emph{Software Quality Journal}, vol.~26, no.~1, pp. 161--191, mar 2018.

\bibitem{Madsen2013PSA}
M.~Madsen, B.~Livshits, and M.~Fanning, ``Practical static analysis of
  javascript applications in the presence of frameworks and libraries,'' in
  \emph{Joint Meeting on Foundations of Software Engineering}, ser. ESEC/FSE
  2013.\hskip 1em plus 0.5em minus 0.4em\relax ACM, 2013, pp. 499--509.

\bibitem{Sun2017AJP}
K.~Sun and S.~Ryu, ``Analysis of javascript programs: Challenges and research
  trends,'' \emph{ACM Comput. Surv.}, vol.~50, no.~4, pp. 59:1--59:34, Aug.
  2017.

\end{thebibliography}

\end{document}